# The Future of Memory: Limits and Opportunities

Samuel Dayo*, Shuhan Liu*, Peijing Li*, Philip Levis*, Subhasish Mitra*,
Thierry Tambe*, David Tennenhouse°, H.-S. Philip Wong*

*Stanford University  °Independent Researcher

## Abstract
Memory latency, bandwidth, capacity, and energy increasingly limit performance. In this paper, we reconsider proposed system architectures that consist of huge (many-terabyte to petabyte scale) memories shared among large numbers of CPUs. We argue two practical engineering challenges, scaling and signaling, limit such designs.

We propose the opposite approach. Rather than create large, shared, homogenous memories, systems explicitly break memory up into smaller slices more tightly coupled with compute elements. Leveraging advances in monolithic/2.5D/3D integration, this compute-memory node provisions private local memory, enabling accesses of node-exclusive data through micrometer-scale distances, and dramatically reduced access cost. In-package memory elements support shared state within a processor, providing far better bandwidth and energy-efficiency than off-package DRAM, which is used as main memory for large working sets and cold data. Hardware making memory capacities and distances explicit allows software to efficiently compose this hierarchy, managing data placement and movement.

## 1 Introduction

The idea of a large, distributed address space of memory is appealing. It allows applications to seamlessly grow beyond a single host while leaving the complexities of caching, consistency, and placement to lower system layers. In the 1980s and 1990s, this idea was explored as distributed shared memory (DSM), informing memory consistency models in modern multi-core and multiprocessor systems.

As memory is increasingly the bottleneck in data center and cloud servers, research is revisiting these ideas to enable a next generation of systems with huge network-attached memories that are pooled, i.e., shared, across many processors. This paper argues that this approach is untenable due to two modern engineering barriers: scaling and signaling. These are practical limits, grounded in physics.

The first, *scaling*, refers to the ability to make transistors and circuits smaller and cheaper with more precise tools and complex manufacturing processes. Memory technology scaling has effectively ended. SRAM's and DRAM's cost per byte have both flattened and there is no roadmap to significantly reduce them in the next five years. As logic continues to shrink, albeit at a slower pace than before, memory becomes a growing fraction of system cost, making it economically and architecturally undesirable to provision large memories. Instead, we should work to improve the efficiency with which memory is utilized.

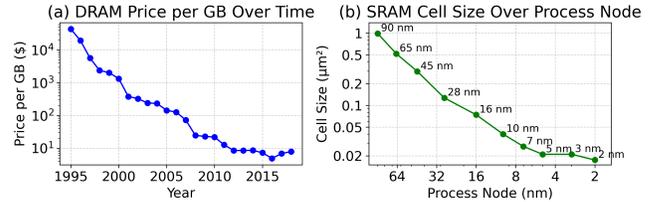

Figure 1: Both DRAM and SRAM have stopped scaling. Reducing cost/byte requires revolutionary changes.

The second barrier, *signaling*, has to do with the energy required to move signals between components at a given bandwidth. It dictates that the energy efficiency and bandwidth to/from memory improve with tighter integration with computational logic.[2] Within a chiplet, access to an SRAM cache line that is far away is slower and/or requires higher energy than a nearby one – while access to one on another chiplet is more expensive still.[1] Going to DRAM across traces on a circuit board is an order of magnitude more expensive; CXL or RDMA to remote memory add even more overheads. These penalties make remote memory prohibitively expensive.

Faced with these barriers, we propose a different approach: physically composable disaggregation. Systems are built from compute-memory nodes that tightly integrate compute with private local memory and on-package shared memory, while using off-package DRAM for bulk capacity. Software explicitly composes the memory system—deciding what data remains local, what is shared across nodes, and what is relegated to DRAM.

## 2 The End of 2D Scaling: SRAM and DRAM

Two-dimensional (2D) semiconductor scaling enabled higher memory density and capacity at reduced cost. However, Figure 1 shows how traditional 2D scaling of both SRAM and DRAM has ended. The cost per byte of DRAM has been flat for over a decade, which is why as servers have scaled up, DRAM has come to dominate system cost. [6] SRAM has reached similar limits: we can no longer make smaller SRAM cells.

For SRAM, the primary constraint stems from transistor dimensions approaching atomic scales: manufacturing tolerances limit transistor matching of the cross-coupled inverter pair, reducing signal margin. Computational logic does not suffer from this issue as each stage restores the digital signal. For DRAM, the primary constraint is the cost of etching the high aspect-ratio capacitors and the complex transistor geometry that guarantees low leakage. More advanced nodes decrease the physical size of DRAM cells, but not the per-memory-cell cost. We can continue to make larger DRAM DIMMs, but their per-byte cost does not decrease.[2]

---

[1]Modern server processors are made up of multiple, separately manufactured *chiplets* that are packaged together into a larger system.

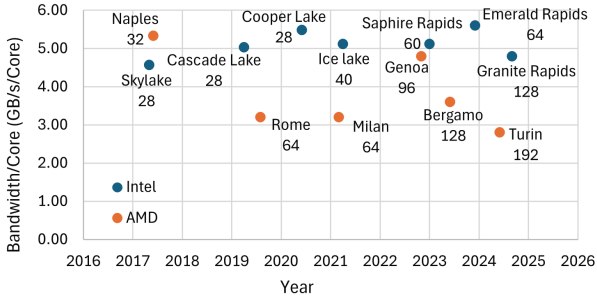

Figure 2: DRAM bandwidth/core for server processors (and core counts). Per-core bandwidth is tagnant.

The primary takeaway from these limits is that enormous memories will be enormously expensive. On-chip caches will not grow faster than chip area, and modern server processors are already huge (AMD SP5 is 5,428mm²). Systems will have to use memory more efficiently.

## 3 Locality = Efficiency AND Bandwidth

Tighter integration increases the bandwidth and energy efficiency of signaling to move data to and from memory. Caches exemplify this principle: L1, L2, and L3 caches all use the same SRAM technology, but L1 caches achieve superior performance through smaller memory bank sizes, finer-grained access, and closer physical proximity to CPU cores.

DRAM bandwidth to a processor socket has been slowly improving: a modern DDR5-5600 DIMM is 358Gbps [4], and the DIMMs per socket grew from 8 to 12, for an aggregate bandwidth of 4.3Tbps. At the same time, however, the number of cores per socket has grown, exceeding or matching bandwidth improvements. Figure 2 shows the per-core bandwidth for Intel and AMD server processor packages since 2018: it has been stagnant.

DRAM's bandwidth limits and energy costs stem from being connected on a printed circuit board (PCB), which has limited number of copper traces and bump pins (e.g. 288 pins for DDR5 [4]). High-bandwidth Memory (HBM) repurposes DRAM dies and moves them closer with improved integration technology. By using an in-package silicon logic base die tucked underneath several DRAM dies, connected by through-silicon-vias, each HBM3E stack has 1024 pins and shorter interconnect distance. This stark difference in pin count directly translates to HBM's bandwidth advantage. Table 1 shows how tighter physical integration allows denser pins, higher bandwidth, and lower energy. Lower pin densities necessitate higher-speed signaling circuits, increasing energy consumption.

These integration limits mean that cores will not see performance improvements from DRAM. Circuit boards cannot accommodate additional DIMMs, and their pin counts are already at their practical limits. Higher signaling speeds across copper traces has a high energy cost.

## 4 Physically Composable Disaggregation

These scaling challenges necessitate a fundamental rethinking of memory hierarchy design – shifting focus from raw capacity to locality, bandwidth, and energy efficiency.

We propose flipping the script on memory "disaggregation" by emphasizing finer-grained integration of memory and compute with increased emphasis on memory utilization—even if it sometimes comes at some modest decrease in compute utilization. At the center of this approach is the compute-memory node, which uses 3D integration technologies to integrate compute with a local memory, stacking memory on top of compute exemplified by AMD's VCache design and Milan-X processors.[7]

Unlike a cache, however, this private local memory is explicitly managed and the exclusive home for node-specific data such as execution stacks and other thread-private state. Accesses over micrometer-scale distances via micro-bumps, hybrid bonds, through-silicon vias, or monolithic wafer-level interconnects, dramatically reduce the latency, energy, and bandwidth bottlenecks of large address spaces. Mirroring practises in modern multi-chiplet processors, shared state that must span nodes–such as locks–is placed in on-package shared memory (e.g., HBM), which, while slower than private local slices, still deliver far better bandwidth and energy-efficiency than off-package DRAM.

However, integration is limited by physical constraints (e.g. thermal dissipation, module size, etc.)[3]. Large memories will continue to require off-package DRAM. Instead of serving as a pooled, flat, shared address space, DRAM becomes a bulk, capacity-driven memory tier for large working sets and cold data, while performance-critical accesses are managed using the faster disaggregated on-package memories. Software *composes* the memory system itself—deciding what data remains local, what is shared, and what is relegated to off-package DRAM—making data placement and movement explicit through abstractions that expose near-zero distance local memory alongside higher-latency shared tiers in a way that enables efficient composition.

| Integration | Pitch | Energy/bit | BW/chip |
|---|---:|---:|---:|
| On-die (5nm) (e.g. SRAM) [1] | 0.028 μm | 5 fJ | 131 TB/s |
| Hybrid bonding (e.g. V-Cache ) [7] | 9 μm | ≈600 fJ | 2.5 TB/s |
| Microbump (e.g. HBM ) [5] | 36 μm | ≈2,000 fJ | 1.2 TB/s |
| C4 solder bump (e.g. DDR) [3] | 730 μm | ≈10,000 fJ | 0.1 TB/s |

Table 1: Four major integration methods: tighter integration has lower energy and higher bandwidth communication.

---

[2]3D DRAM promises to improve bit density but its cost is unknown.

[3]E.g., one literally cannot fit terabytes of DRAM inside a chip package.

# References

[1] Tsung-Yung Jonathan Chang, Yen-Huei Chen, Wei-Min Chan, Hank Cheng, Po-Sheng Wang, Yangsyu Lin, Hidehiro Fujiwara, Robin Lee, Hung-Jen Liao, Ping-Wei Wang, and others. 2020. A 5-nm 135-mb SRAM in EUV and high-mobility channel FinFET technology with metal coupling and charge-sharing write-assist circuitry schemes for high-density and low-V MIN applications. *IEEE Journal of Solid-State Circuits* 56, 1 (2020), 179–187.

[2] R. Ho, K.W. Mai, and M.A. Horowitz. 2001. The future of wires. *Proceedings of the IEEE* 89, 4 (2001), 490–504. https://doi.org/10.1109/5.920580

[3] Micron. 2025. Micron GDDR7 Memory Product Brief. Retrieved from https://assets.micron.com/adobe/assets/urn:aaid:aem:087330f6-6d71-4575-b622-cb10f20cdaf0/renditions/original/as/gddr7-product-brief.pdf

[4] Micron. 2025. DDR5 DRAM. Retrieved from https://www.micron.com/products/memory/dram-components/ddr5-sdram?srsltid=AfmBOorJp4ijYawwQ-dMiW7YaljuBZ9Ws66BkjLNzNj_3W8mFDCZZC89

[5] Raj Narasimhan. 2025. Micron continues memory leadership with HBM3E 36GB 12-high. Retrieved from https://www.micron.com/about/blog/applications/ai/micron-continues-memory-industry-leadership-with-hbm3e-12-high-36gb

[6] Neel Patel, Amin Mamandipoor, Derrick Quinn, and Mohammad Alian. 2023. XFM: Accelerated Software-Defined Far Memory. In *(MICRO '23)*, 2023.

[7] John Wuu, Rahul Agarwal, Michael Ciraula, Carl Dietz, Brett Johnson, Dave Johnson, Russell Schreiber, Raja Swaminathan, Will Walker, and Samuel Naffziger. 2022. 3D V-Cache: the Implementation of a Hybrid-Bonded 64MB Stacked Cache for a 7nm x86-64 CPU. In *Proceedings of ISSCC*, 2022.